\documentclass[a4paper,12pt]{article}

\usepackage{amsmath}
\usepackage{amssymb}
\usepackage{authblk}
\usepackage{geometry}
\usepackage{graphicx}

\begin{document}

\title{
Percolate or die: Multi-percolation decides the struggle \\
between competing innovations
}

\author[1,*]{Carlos P.\ Roca}
\author[2]{Moez Draief}
\author[1,3]{Dirk Helbing}

\affil[1]{Chair of Sociology, in particular of Modeling and Simulation, 
ETH Zurich, Clausiusstrasse 50, 8092 Zurich, Switzerland}
\affil[2]{Intelligent Systems and Networks Group, Imperial College London, 
South Kensington Campus, London SW7 2AZ, United Kingdom}
\affil[3]{Santa Fe Institute, 1399 Hyde Park Road, Santa Fe, 
New Mexico 87501, USA}
\affil[*]{e-mail: cproca@ethz.ch}

\date{\today}

\maketitle

\begin{abstract}

\bfseries

Competition is one of the most fundamental phenomena in physics, biology and 
economics.
Recent studies of the competition between innovations have highlighted the 
influence of switching costs and interaction networks, but the problem is 
still puzzling. 
We introduce a model that reveals a novel multi-percolation process, which 
governs the struggle of innovations trying to penetrate a market.
We find that innovations thrive as long as they percolate in a population, 
and one becomes dominant when it is the only one that percolates. 
Besides offering a theoretical framework to understand the diffusion of 
competing innovations in social networks, our results are also relevant to 
model other problems such as opinion formation, political polarization, 
survival of languages and the spread of health behavior.

\end{abstract}

The analysis of percolation in random media has become a very popular 
framework over the last decades to address a wide variety of phenomena 
in disordered systems, such as 
oil mining in porous reservoirs, fire spreading in forests, fracture patterns 
in rocks, electromagnetic properties of composite materials, etc \cite{sahimi:1994:applications}. 
More recently, it has also been applied to shed light on social phenomena, 
namely the diffusion of opinions \cite{shao:2009:dynamic} and innovations 
\cite{goldenberg:2000:marketing} in social networks.
All of the aforementioned systems can be modeled as percolation problems. 
More precisely, they can be abstracted as a network of nodes representing 
the topology of the random medium, wherein nodes can be either 
``empty'' or ``occupied'', depending on the state of their neighbors.
Starting from an initial condition where some nodes are occupied, an occupied 
node becomes empty if the number of its occupied neighbors goes below a threshold 
$k$, the index of the percolation process 
($k=2$ for standard percolation \cite{stauffer:1991:introduction} and $k \ge 3$ 
for bootstrap or $k$-core percolation 
\cite{chalupa:1979:bootstrap,dorogovtsev:2006:k-core}).
The underlying switching dynamics is therefore assumed to be \emph{unidirectional}. 

Here we introduce a percolation model that generalizes this powerful theoretical approach. 
Our extension assumes that nodes are of two types A or B, and that a node changes 
type when the number of neighbors of the same type is less than $k$. 
Consequently both changes A-to-B and B-to-A are possible, i.e.\ we are considering a 
\emph{bi-directional} percolation dynamics instead. 
Figure 1 provides an example which illustrates the fundamental difference 
between both percolation processes.

The problem we want to address is the competition between innovations 
\cite{arthur:1989:competing}. 
Competition between products, tools or technical standards is ubiquitous. 
Well-known examples are railway gauges, keyboard layouts, computer operating 
systems, high-definition video standards, e-book readers, etc.
The reasons that determine the outcome of these fierce competitions have puzzled researchers of different disciplines for a long time 
\cite{rogers:2003:diffusion,amini:2009:marketing}. 
Previous work has highlighted the combined influence of intrinsic benefits of 
each option together with costs incurred due to switching 
\cite{klemperer:1995:competition}. In addition, it has been 
suggested that social structure, i.e.\ the network of social relationships 
in a group or population, would play a crucial role 
\cite{granovetter:1978:threshold}. 
So far, however, there has been little analytical work that elucidates the 
outcome of such competitions. 
In this work we show that the competition between innovations can be 
understood as a bi-directional percolation process, which ultimately determines 
the fate of the options in contest. 

To start with, let us consider a simple model with two competing options, 
A and B (for example Blu-ray Disc vs HD DVD), whose benefits to individuals 
depend on intrinsic factors as well as on the acceptance by others in a certain 
social neighborhood. 
This can be modeled as a coordination problem \cite{myerson:1991:game,skyrms:2003:stag}, 
in which individuals choosing one of the two options A or B obtain a payoff 
$\pi_\mathrm{A} = q \tilde{x}_\mathrm{A}$ and 
$\pi_\mathrm{B} = (1-q) \tilde{x}_\mathrm{B}$ respectively.
The relative advantage of one option over the other is represented 
by the parameter $q$, where $0 \le q \le 1$.
Quantities $\tilde{x}_\mathrm{A}$ and $\tilde{x}_\mathrm{B}$ give, respectively, 
the proportion of people adhering to option A or B among the social acquaintances 
who have an influence on the individual's decision, such us family members, 
friends, co-workers, etc 
($\tilde{x}_\mathrm{A} + \tilde{x}_\mathrm{B} = 1$ for every individual). 
In addition, we consider that changing option entails some switching cost, 
which is called $c_\mathrm{A}$ for a follower of option A who changes to B, 
and $c_\mathrm{B}$ in the opposite case. 
Thus, A- and B-individuals have the following effective payoff 
matrices
\begin{equation}
\label{eq:payoffs}
\mathrm{A:}\;
\begin{pmatrix}
q & 0 \\
-c_\mathrm{A} & 1-q - c_\mathrm{A}
\end{pmatrix}
\qquad\qquad
\mathrm{B:}\;
\begin{pmatrix}
q - c_\mathrm{B} & -c_\mathrm{B} \\
0 & 1-q
\end{pmatrix} ,
\end{equation}
where we follow the standard convention for symmetric games: 
rows represent own strategy and columns that of the interaction partner.

For the moment, we assume that individuals are able to assess 
to a good extent the benefits and drawbacks of options A and B, 
and also the degree of penetration of each option in their social neighborhood, 
i.e.\ we assume a low level of uncertainty in the decision-making process 
(more on this important point later). 
Therefore, individuals choose a best response to the current state of their 
social context according to the payoffs expected from Eq.~(\ref{eq:payoffs}). 
As a consequence, A-individuals change to option B if the proportion of A-neighbors 
is below a certain threshold, namely $\tilde{x}_\mathrm{A} < 1 - q - c_\mathrm{A}$, 
while B-individuals switch if the proportion of B-neighbors is less than certain 
value, $\tilde{x}_\mathrm{B} < q - c_\mathrm{B}$.

This defines an evolutionary game \cite{nowak:1992:evolutionary,gintis:2009:game}, 
which consists in a dynamical system with state variable $x$, the global density of 
the followers of one of the options. 
We set $x=x_\mathrm{B}$ without loss of generality. 
Disregarding the effect of social structure for the moment, the evolution 
of $x_\mathrm{B}$ can easily be calculated assuming a well-mixed population 
\cite{hofbauer:1998:evolutionary}, equivalent to the mean field hypothesis in 
physics or the representative agent in economics. 
It posits that every individual has, in her social neighborhood, a proportion of 
A- or B-individuals equal to the respective global densities, i.e.\ for every 
individual and at any time $\tilde{x}_\mathrm{B} = x_\mathrm{B}$. 
Under this assumption, the population rapidly reaches an equilibrium with stationary 
value $x_\mathrm{B}^* = \lim_{t\to\infty}x_\mathrm{B}(t)$
\begin{equation}
\label{eq:well-mixed}
x_\mathrm{B}^* =
\begin{cases}
0, & \mathrm{if}\; x_\mathrm{B}^0 < q - c_\mathrm{B}, \\
x_\mathrm{B}^0, & \mathrm{if}\; 
  q - c_\mathrm{B} \le x_\mathrm{B}^0 \le q+c_\mathrm{A}, \\
1, & \mathrm{if}\; x_\mathrm{B}^0 > q + c_\mathrm{A} ,
\end{cases}
\end{equation}
where $x_\mathrm{B}^0$ represents the initial density of individuals following 
option B. Equation~(\ref{eq:well-mixed}) shows that under well-mixed 
conditions switching costs induce the appearance of a heterogeneous 
state, in which both competing options keep a share of the population. 
If costs are left out ($c_\mathrm{A}=c_\mathrm{B}=0$), then we find the standard 
solution of a coordination game, with an unstable equilibrium at 
$x_\mathrm{B}^* = x_\mathrm{B}^0$, which separates the basins of attraction of 
the stable equilibria $x_\mathrm{B}^* = 0$ and $x_\mathrm{B}^* = 1$ 
\cite{helbing:1992:mathematical}.

Let us now consider this model embedded in a social network 
\cite{wasserman:1994:social,vega-redondo:2007:complex}, that is, 
in a network of social relationships which determines who interacts with whom 
\cite{nowak:1992:evolutionary,nakamaru:2004:spread,galeotti:2010:network}. 
Here we use regular random networks \cite{newman:2002:random}, which are networks 
where nodes are linked at random and where each node has the same number of 
neighbors, or degree, $z$. 
Such networks are known to have the most neutral effect on evolutionary games 
\cite{roca:2009:evolutionary}, just fixing neighborhood size and preserving 
the social context of individuals. 
They avoid particular effects that some topological features, such as clustering or degree heterogeneity, may have \cite{szabo:2007:evolutionary}, which could obscure 
the processes we want to reveal here. 

\begin{figure}
\centering
\includegraphics[width=0.8\textwidth]{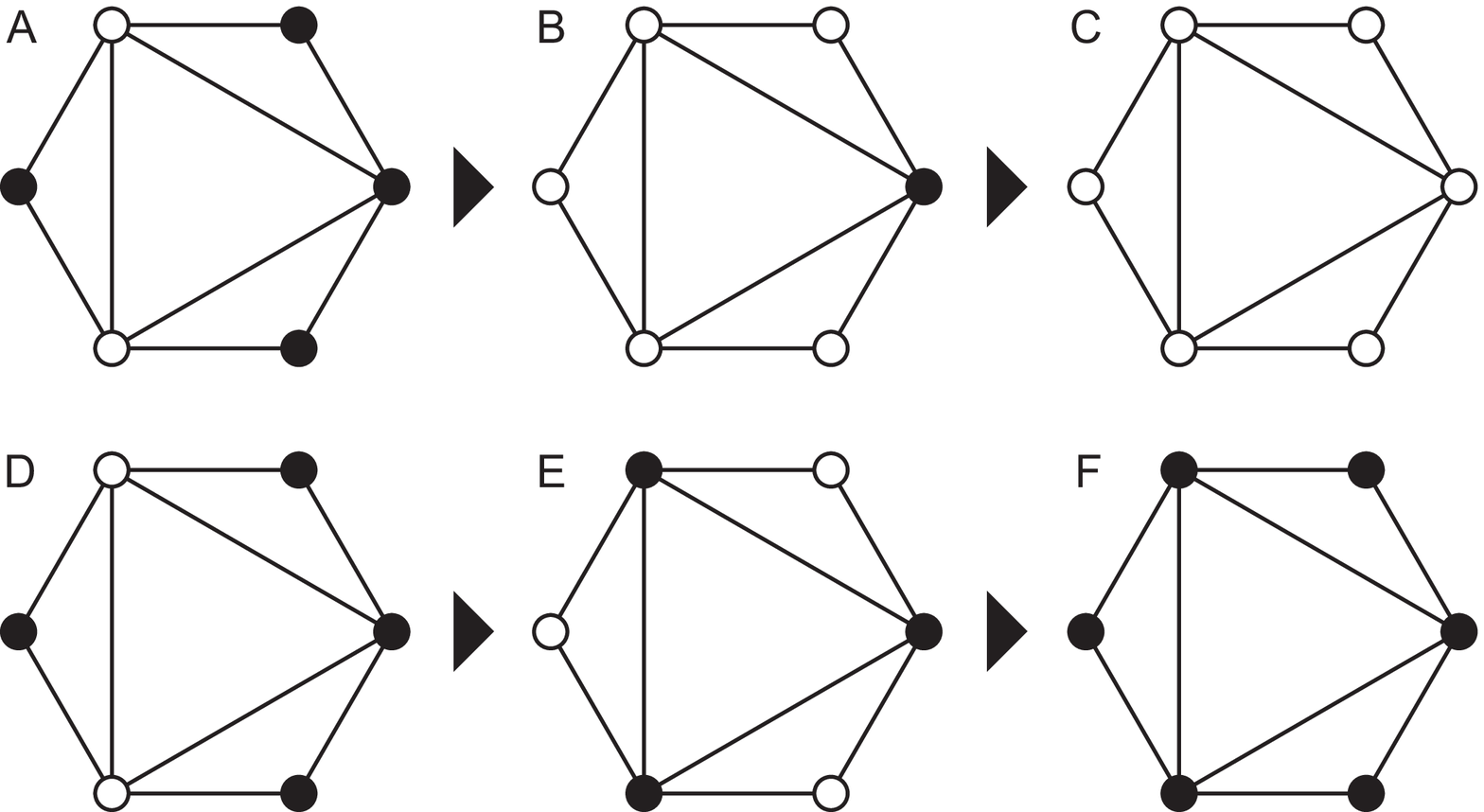}
\caption{Comparison of unidirectional (panels A--C) vs bi-directional 
percolation (panels D--F). 
In unidirectional percolation, occupied nodes (in black) become empty (in white)
when they have less than $k$ occupied neighbors (in this example $k=2$). 
In the end there is no occupied node that survives the percolation pruning. 
With bi-directional percolation, both white and black nodes switch color when they 
have less than $k=2$ neighbors of their same color.
The end result in this case is an all-black graph, with no white nodes surviving 
the competitive bi-directional percolation process.
All black nodes end up with two black neighbors and they are connected, 
hence they form a percolating cluster. 
Notice that although both cases have the same initial condition and percolation index 
$k$, the outcome is opposite.}
\label{fig:01}
\end{figure}

\begin{figure}
\centering
\includegraphics[width=0.8\textwidth]{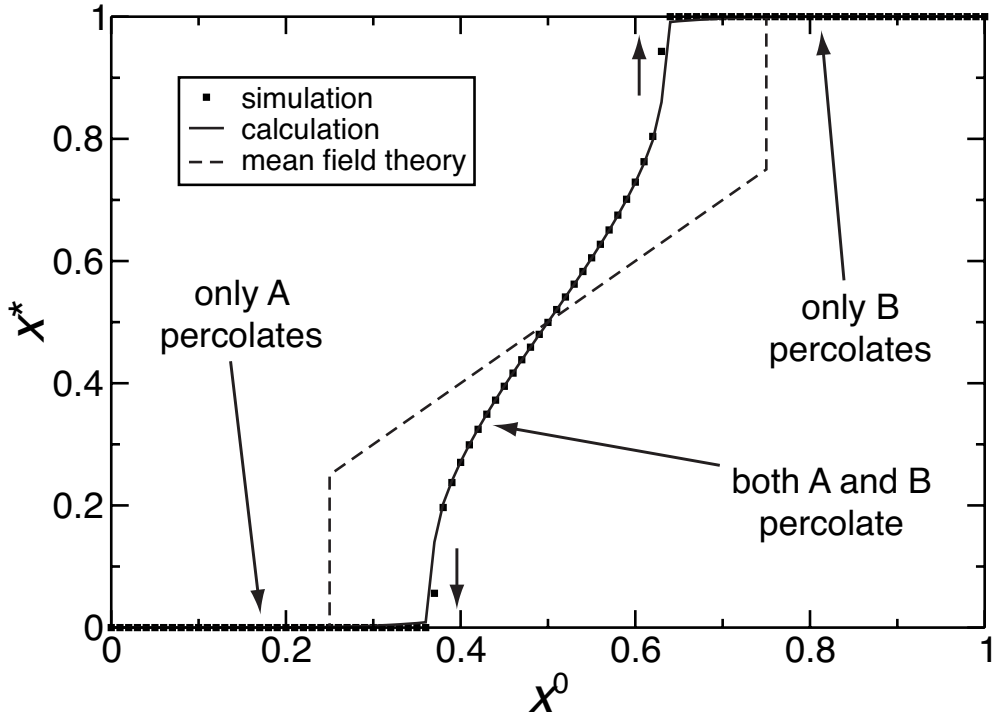}
\caption{The fate of two options in contest is determined by the underlying process 
of multi-percolation, taking place on the social network.  
The graph shows the stationary density of B-individuals $x^*$ as a function of 
the initial density $x^0$ (simulation results, black squares). 
Parameter $q=0.5$, so both options A and B are intrinsically equally good.
Switching costs are also symmetric, with values $c_\mathrm{A}=c_\mathrm{B}=0.25$. 
As a result, both percolation indices have the same value 
$k_\mathrm{A} = k_\mathrm{B} = 3$. 
Interactions take place in a regular random network of degree $z=9$. 
The difference with the prediction of mean theory (dashed line) demonstrates 
the crucial role played by the social network. 
Labels indicate the possible regions of behavior, depending on the percolation 
of one option, the other or both.
Notice that a heterogeneous population is sustainable only when both options 
percolate, but this case occurs for a significant range of initial conditions. 
Small arrows near abscissa axes mark the critical density to attain percolation 
of each strategy, as predicted by a calculation based on standard unidirectional 
percolation. 
The discrepancy with simulation results highlights the fact that the mutual 
interference between both percolation processes changes the percolation thresholds. 
This is confirmed by an analytical calculation that takes into account this interplay 
(solid line).}
\label{fig:02}
\end{figure}

Figure~\ref{fig:02} displays simulation results for this model, showing the 
stationary density of B-individuals $x_\mathrm{B}^*$ as a function of their 
initial density $x_\mathrm{B}^0$ (see the Materials and Methods section for full details about the 
simulations).
Notably, there are large differences to mean field theory, which predicts 
a heterogeneous population for a much wider range of initial conditions. 
In order to understand this deviation we have to consider the time evolution of the 
model. 
To that purpose, it is better to start with a simpler case, setting one of 
the switching costs with so large a value that it prevents the switching of 
individuals following that strategy. 
For example, let us set $c_\mathrm{A} \ge 1-q$, so that only B-individuals can 
change to option A. 
This switching takes place when the proportion of B-individuals in the neighborhood 
of the focal individual satisfies $\tilde{x}_\mathrm{B} < q - c_\mathrm{B}$. 
Hence the subsequent dynamics exactly coincides with the pruning of nodes of a 
standard site percolation process with unidirectional dynamics (see 
Fig.~\ref{fig:01}), A- and B-individuals corresponding to empty and occupied 
nodes, respectively.
When B-nodes become A-nodes, they leave other B-nodes with fewer B-neighbors. 
The process repeats until it stabilizes to a subset of B-nodes, 
all of which have $(q - c_\mathrm{B})$ or more B-neighbors. 
When the size of this subset is a non-negligible fraction of the size of the full 
graph, or infinite in the case of infinite graphs, then percolation is said to occur 
\cite{stauffer:1991:introduction}. 
The appearance of a percolating cluster constitutes a phase transition, and it 
takes place when the initial density of occupied nodes is larger than a critical 
density. 
In our case, the index of the percolation process of B-individuals switching 
to option A, called $k_\mathrm{B}$, is given by
\begin{equation}
\label{eq:kb}
k_\mathrm{B} = \lceil z (q-c_\mathrm{B}) \rceil .
\end{equation}
Herein, $\lceil x \rceil$ denotes the smallest integer equal or larger than $x$.
Conversely, considering only the transitions of A-individuals to option B, 
we have \emph{another} percolation process with index $k_\mathrm{A}$, whose value 
is given by
\begin{equation}
\label{eq:ka}
k_\mathrm{A} = \lceil z (1-q-c_\mathrm{A}) \rceil .
\end{equation}
Note that $k=0$ and $k=1$ are degenerate cases, whereas $k=2$ is the index of 
standard percolation and $k \ge 3$ corresponds to bootstrap or $k$-core percolation. 

The actual dynamics of our model is given by the competition between these two percolation processes. The dynamics is therefore bi-directional, with both 
transitions A-to-B and B-to-A taking place simultaneously. 
A calculation based on standard unidirectional percolation, applied to 
each process separately, estimates the percolation thresholds only poorly, 
as the arrows in Fig.~\ref{fig:02} show. 
It is also possible, however, to take into account the interference between 
both processes, with a recursive calculation on the switching times of 
individuals. 
Figure~\ref{fig:02} shows the excellent agreement of this calculation with 
the computer simulations, and it clearly demonstrates that mutual interference 
between both percolation processes occurs.
Interestingly, we find that this interplay supports the success of the dominated option, i.e.\ it allows some individuals following the minor option 
to percolate with initial conditions for which unidirectional percolation 
does not occur 
(range $0.36 \lesssim x_\mathrm{B}^0 \lesssim 0.38$ for option B, 
and range $0.62 \lesssim x_\mathrm{B}^0 \lesssim 0.64$ for option A). 
Individuals who have switched obviously promote percolation of the newly acquired 
option, as the switching increases the density of that option in the neighborhoods 
of adjacent nodes. 
The time scale of switching for the major option is much faster than for the 
minor one \cite{grimmett:1999:percolation}. 
This implies that the pruning of nodes for the major option is virtually done by 
the time the pruning of the minor option proceeds, with the consequence 
that only changes of major to minor option have time to effectively foster percolation 
of the latter option.
More importantly, this analytical theory confirms that the competition 
between options A and B gives rise to a bi-directional percolation process, which 
allows a simple rationale for the outcome: 
\emph{In the context of competing innovations, percolation of an option means 
survival and, as long as it is the only one that percolates, it also implies 
dominance of the market.}
We refer the reader to Supporting Information for details of the analytical 
calculations and further discussion.

\begin{figure}
\centering
\includegraphics[width=0.8\textwidth]{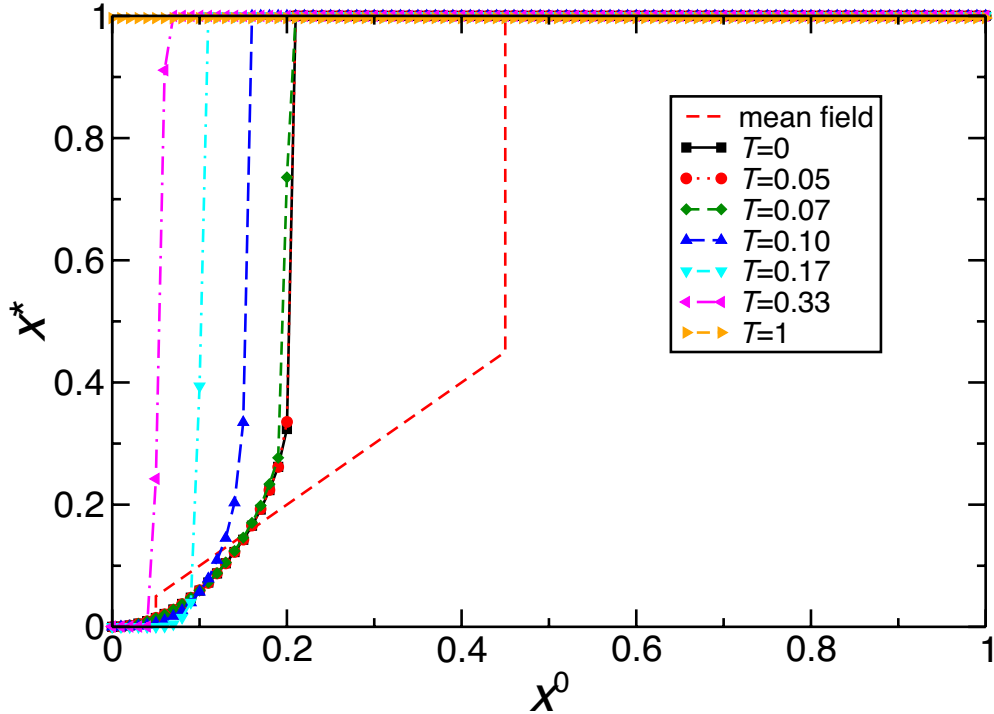}
\caption{Noise only has an effect on the multi-percolation process when the amount 
is large.
The graph shows the stationary density of B-individuals $x^*$ as a function of the 
initial density $x^0$, for different amounts of uncertainty or noise in the decision 
rule of individuals (simulation results, see lines, colors and symbols in legend).
Compared to Fig.~\ref{fig:02}, in this case $q=0.25$, so option B is superior 
to option A.
Switching costs are equal for both options, with values 
$c_\mathrm{A} = c_\mathrm{B} = 0.2$.
The social network is a regular random network of degree $z=6$. 
For $T=0$ (no noise), the asymmetry in intrinsic value between the options translates into different percolation indices, $k_\mathrm{A}=4$ and $k_\mathrm{B}=1$, 
which causes different kinds of transitions to homogeneous population (A- or B-dominated).
This fact favors option B compared to the mean field prediction.
Additional curves show results for non-zero amounts of noise. 
Moderate noise does not change the result qualitatively and, strikingly, 
larger amounts reinforce the superior option B rather than yielding an more balanced 
outcome.}
\label{fig:03}
\end{figure}

The joint influence of switching costs and social networks becomes most 
salient when one of the options is intrinsically superior to the other, 
i.e.\ when $q \ne 0.5$. 
Figure~\ref{fig:03} shows an example, displaying again the asymptotic density 
of B-individuals $x_\mathrm{B}^*$ as a function of their initial density 
$x_\mathrm{B}^0$ (see solid line and black squares). 
In this case, the asymmetry of the game results in different percolation indices 
for each option, namely $k_\mathrm{A}=4$ and $k_\mathrm{B}=1$ (see 
Eqs.~(\ref{eq:kb})--(\ref{eq:ka})), which causes a continuous transition towards an 
A-dominated population ($x_\mathrm{B}^0 \lesssim 0.1$), but a discontinuous one in 
the B-dominated case ($x_\mathrm{B}^0 \gtrsim 0.2$). 
The difference between both transitions \cite{achlioptas:2009:explosive} 
originates from the characteristic transition of standard percolation, in the 
former case, versus that of bootstrap percolation, in the latter. 
Interestingly, the net effect of this imbalance between the two 
competing percolation processes is a fostering of the superior option B. 
Note that if the same game, with or without switching costs, took place on 
a well-mixed population, a symmetric outcome around $x_\mathrm{B}^0 = q = 0.25$ 
would result instead (see Eq.~(\ref{eq:well-mixed}) and dashed red line in 
Fig.~\ref{fig:03}).

Let us finally address the issue of uncertainty or noise in the decision rule of 
individuals. 
To this end, we assume that individuals choose options stochastically, with a 
probability that follows a multi-nomial logit model 
\cite{mcfadden:1974:conditional,goeree:1999:stochastic}, 
which is also known in physics as Fermi rule 
\cite{blume:1993:statistical,traulsen:2006:stochastic}. 
Specifically, if the expected variation in payoff resulting from a change of 
option is $\Delta\pi$, then the probability of switching strategy is assumed 
to be $1/(1 + \exp(- \Delta\pi / T))$. 
The parameter $T$ determines the amount of noise in the decision process. 
In the limit $T \to 0$, noise disappears and we have the deterministic dynamics 
used so far.
Additional curves in Fig.~\ref{fig:03} display the influence of noise, 
showing that the qualitative behavior of the model remains the same for low 
to moderate amounts of noise. 
It is striking, however, that the evolution of the population is biased towards 
the superior option B, i.e.\ noise reinforces the imbalance between options 
rather than washing it out. 

In conclusion, we have shown that the competition between several options 
gives rise to bi-directional (or, more generally, multi-directional) 
percolation processes which can be analytically understood. Multi-percolation 
thus provides a powerful theoretical tool to understand the problem of competition between innovations in networks. 
It offers predictions about the survival and dominance of the options in 
contest, as well as insights into their dynamical evolution in time. 
Our general finding is that percolation of an option implies its survival and, 
if only one option percolates, it will be eventually supported by everyone. 
The latter may be favorable, for example when it promotes a shared technical standard. Nevertheless, it could also create monopolies or endanger pluralism. 
Our conclusions are expected to be also relevant to the results of marketing and 
political campaigns, and to the diffusion of opinions 
\cite{granovetter:1978:threshold} and behavior \cite{centola:2010:spread}. 
Model variants or extensions may also describe the spread of health behavior, such as 
obesity, smoking, depression or happiness \cite{smith:2008:social}.
We recognize that the practical applicability of this theory requires a good 
knowledge of the social network and a sound modeling of the decision making process of 
individuals \cite{traulsen:2010:human}, 
but with the emergent availability of massive social data and information 
technologies, we envisage that this goal will be attainable soon.

\section*{Methods Summary: Computer Simulations}

All the simulation results reported in the main text have been 
obtained according to the procedures described in the following.
Population size is $10^4$ individuals. 
Regular random networks are generated by randomly assigning links between nodes, 
ensuring that each node ends up with exactly the same number of links.
Results have been obtained with synchronous update. This means that, first, 
the next strategy is calculated for all individuals and, then, it is updated for them 
all at once. We have also checked the influence of asynchronous update, which assigns 
next strategies to individuals proceeding one by one in random order, but we have 
found no significant qualitative difference in the results. 
The times of convergence allowed for the population to reach a stationary state are 
$10^2$ steps for the main model (best response rule) and $10^3$ steps for the model with 
noise (multi-nomial logit or Fermi decision rule). 
We have verified that these are sufficiently large values. 
In the first case, population reaches a frozen state and, in the second one, results 
do not change if a time of convergence $10^4$ steps is used instead.
Results plotted in the graphs correspond, for each data point, to an average of 100 realizations. 
Each realization is carried out with a newly generated random network, where 
strategies were randomly assigned to individuals in accordance with the initial densities of both strategies.

\section*{Acknowledgments}
C.\ P.\ R.\ and D.\ H.\ were partially supported by the Future and Emerging 
Technologies programme FP7-COSI-ICT of the European Commission through 
project QLectives (grant no. 231200).

\linespread{1.1}

\clearpage

\section*{Supplementary Information: Analytical Theory}

We study a coordination a game with switching costs. A-individuals have the following 
payoff matrix 
\begin{equation}
\label{si:eq:payoffs-A}
\begin{pmatrix}
q & 0 \\
-c_\mathrm{A} & 1-q - c_\mathrm{A}
\end{pmatrix} ,
\end{equation}
while B-individuals have
\begin{equation}
\label{si:eq:payoffs-B}
\begin{pmatrix}
q - c_\mathrm{B} & -c_\mathrm{B} \\
0 & 1-q
\end{pmatrix} .
\end{equation}
All three parameters verify $q, c_\mathrm{A}, c_\mathrm{B} \in [0,1]$.

We call $x^t$ the global density of B-individuals in the population at time step $t$, 
$x^0$ its initial value and $x^*$ its asymptotic value ($x^* = \lim_{t\to\infty}x^t$). 
At time $t=0$ individuals are randomly assigned options A or B, with probabilities 
$x_\mathrm{A}^0=(1-x^0)$ or $x_\mathrm{B}^0=x^0$ respectively.
Given an individual $i$, we call $\tilde{x}_i$ the proportion of B-individuals in 
her neighborhood. 
Then, an A-individual $i$ changes to option B when her neighborhood verifies
\begin{equation}
\label{si:eq:switch-A}
\tilde{x}_i > q + c_\mathrm{A} ,
\end{equation}
whereas a B-individual $j$ switches strategy when
\begin{equation}
\label{si:eq:switch-B}
\tilde{x}_j < q - c_\mathrm{B} .
\end{equation}

The above strategy switchings correspond to the pruning of nodes of two 
competing site percolation processes, whose indices are respectively 
\begin{equation}
\label{si:eq:percolation-A}
k_\mathrm{A} = \lceil z (1-q-c_\mathrm{A}) \rceil 
\end{equation}
and
\begin{equation}
\label{si:eq:percolation-B}
k_\mathrm{B} = \lceil z (q-c_\mathrm{B}) \rceil ,
\end{equation}
wherein $\lceil x \rceil$ denotes the smallest integer equal or larger than $x$.

In the pruning of nodes associated with a site percolation process, the nodes that 
remain in the stationary state are those which have $k$ or more neighbors of the 
same strategy and hence they belong to a percolating cluster of that strategy. 
We call $p_\mathrm{A}$ and $p_\mathrm{B}$ the probabilities that a node belongs 
to a percolating cluster assuming an \emph{independent} percolation process with  transitions $\mathrm{A}\to\mathrm{B}$ and $\mathrm{B}\to\mathrm{A}$ respectively.
Obviously, $p_\mathrm{A} \le x_\mathrm{A} = 1-x^0$ and 
$p_\mathrm{B} \le x_\mathrm{B} = x^0$. 
Both probabilities can be calculated for infinite Bethe lattices 
\cite{chalupa:1979:bootstrap,dorogovtsev:2006:k-core}. 
They offer a rough approximation of the behavior of the model, assuming no 
interference between both processes, which yields the following prediction for the 
asymptotic density of B-individuals
\begin{equation}
\label{si:eq:approx-no-ifc}
x^* = 1 - x^0 + p_\mathrm{B} - p_\mathrm{A} .
\end{equation}

This approximation obviously fails to account for the change in the percolation probabilities that the simulation results reflect. 
Therefore, we propose the following analytical theory, which takes into 
account the interplay between both competing percolation processes. 

Let us first define a simplified version of the model, which we call the 1T-model. 
This model is the same as the original one, with the only difference that each node 
can only switch strategy once. That is, nodes change strategy according to 
Eqs.~(\ref{si:eq:switch-A})--(\ref{si:eq:switch-B}), but once they have changed, 
they stick to the new strategy, no matter what Eqs.~(\ref{si:eq:switch-A})--(\ref{si:eq:switch-B}) dictate. 
Note that this idea can be generalized to a $n$T-model, where each node is allowed 
to switch at most $n$ times. 
Our original model could also be called $\infty$T-model, i.e.\ when $n$ is 
unbounded.

\begin{figure}
\centering
\includegraphics[width=0.8\textwidth]{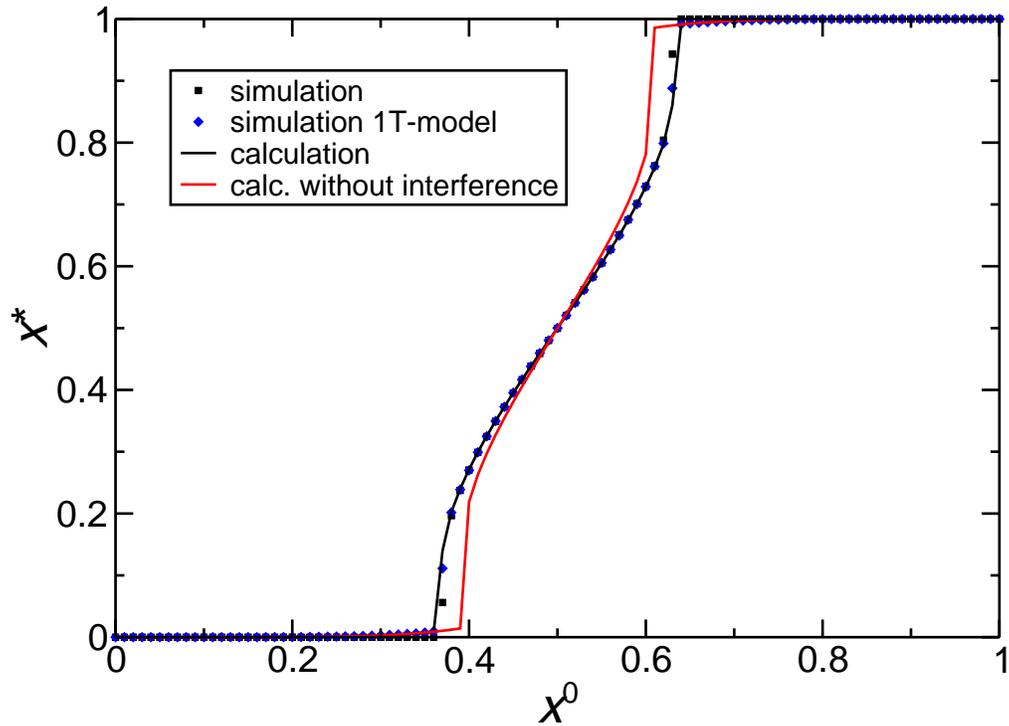}
\caption{Asymptotic density of B-nodes $x^*$ as a function of the initial density 
$x^0$. The model parameters are the same as in Fig.~2 of the main text: game $q=0.5$, 
costs $c_\mathrm{A}=c_\mathrm{B}=0.25$, network degree $z=9$. 
The corresponding indices of the percolation processes are 
$k_\mathrm{A}=k_\mathrm{B}=3$.}
\label{si:fig:01}
\end{figure}

\begin{figure}
\centering
\includegraphics[width=0.8\textwidth]{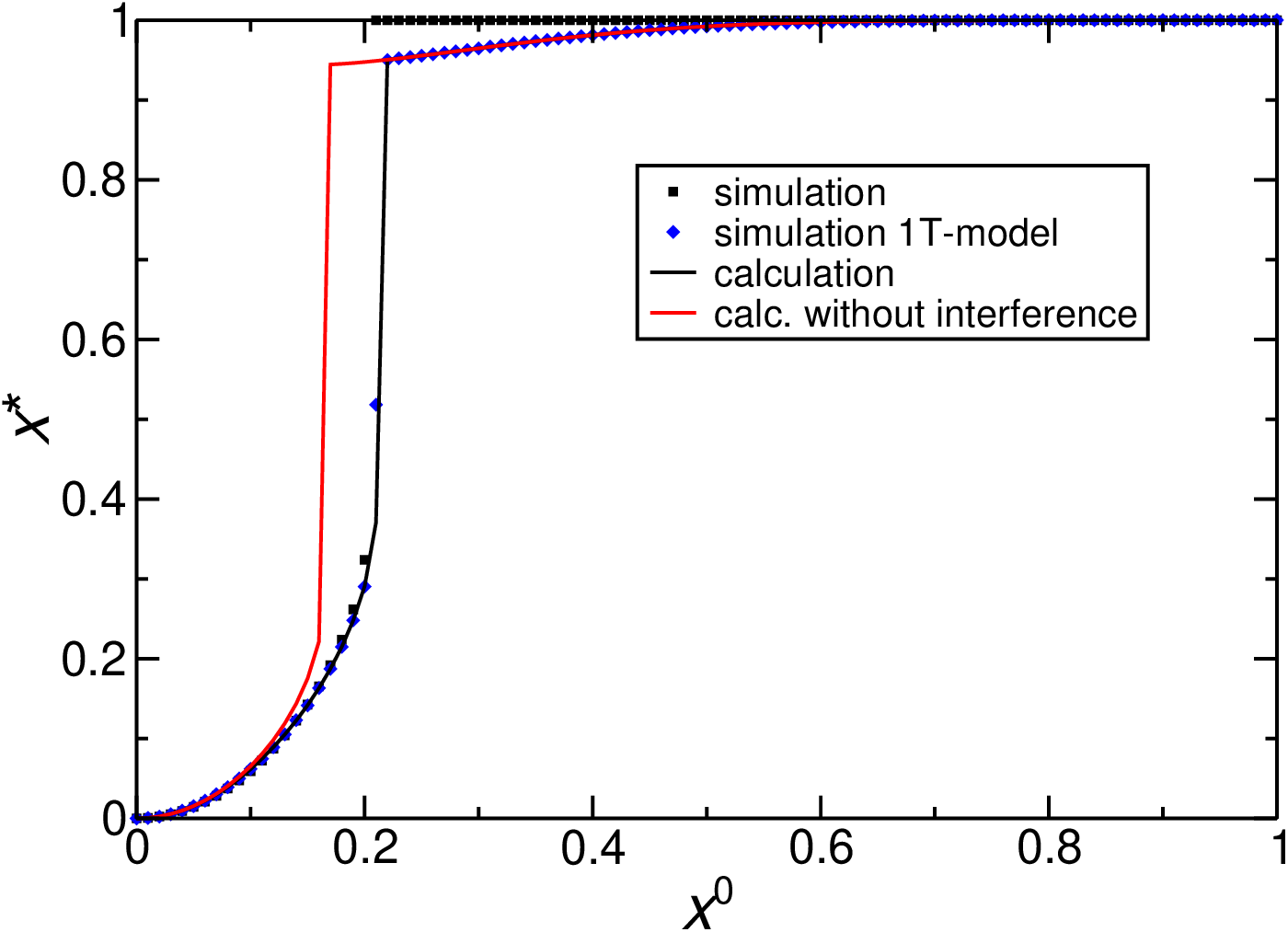}
\caption{Asymptotic density of B-nodes $x^*$ as a function of the initial density 
$x^0$. The model parameters are the same as in Fig.~3 of the main text: game $q=0.25$, 
costs $c_\mathrm{A}=c_\mathrm{B}=0.2$, network degree $z=6$. 
The corresponding indices of the percolation processes are 
$k_\mathrm{A}=4$, $k_\mathrm{B}=1$.}
\label{si:fig:02}
\end{figure}

\begin{figure}
\centering
\includegraphics[width=0.8\textwidth]{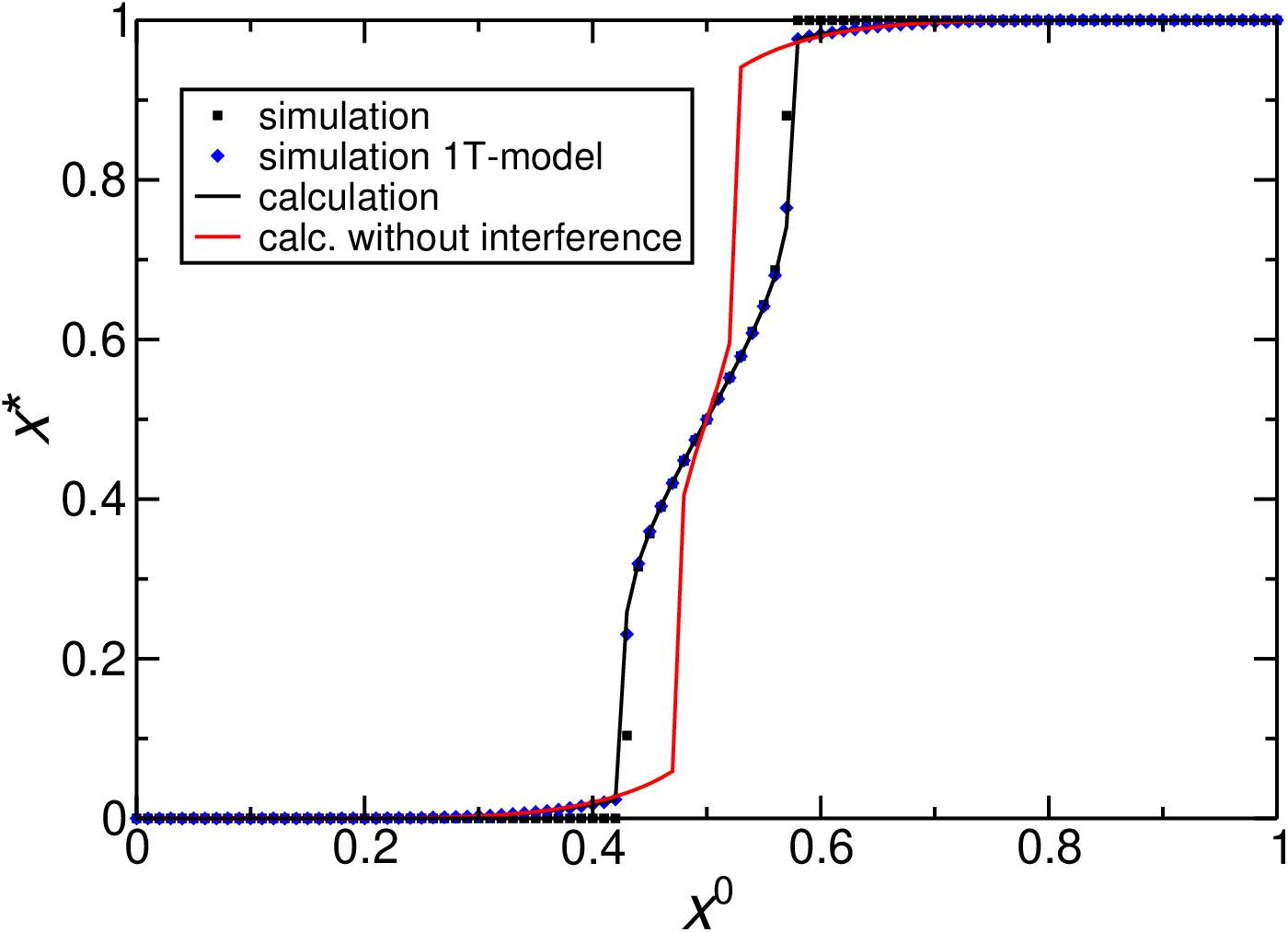}
\caption{Asymptotic density of B-nodes $x^*$ as a function of the initial density 
$x^0$. Model parameters are: game $q=0.5$, costs $c_\mathrm{A}=c_\mathrm{B}=0.2$, 
network degree $z=11$. 
The indices of the percolation processes are 
$k_\mathrm{A}=k_\mathrm{B}=4$.}
\label{si:fig:03}
\end{figure}

\begin{figure}
\centering
\includegraphics[width=0.8\textwidth]{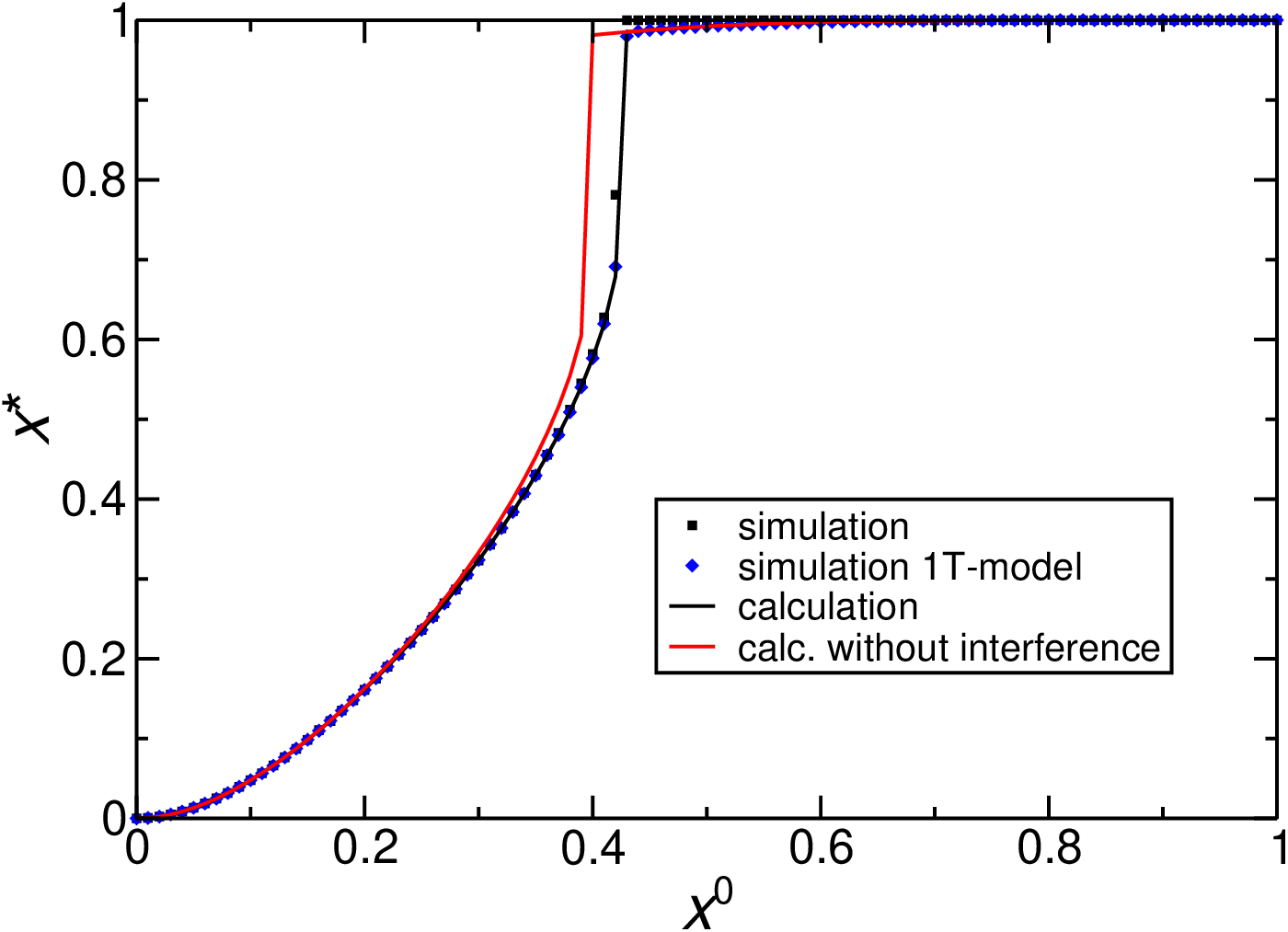}
\caption{Asymptotic density of B-nodes $x^*$ as a function of the initial density 
$x^0$. Models parameters are: game $q=0.3$, costs $c_\mathrm{A}=c_\mathrm{B}=0.2$, 
network degree $z=6$. 
The indices of the percolation processes are 
$k_\mathrm{A}=3$, $k_\mathrm{B}=1$.}
\label{si:fig:04}
\end{figure}

In fact, we can assume in the original model that most nodes will switch only once 
or never. 
For a node to switch, it is required that the number of neighbors of the same 
strategy is below a given threshold $k$. 
Once it has switched, it will have a number of neighbors of the (new) strategy 
larger than $z-k$ ($z$ is the degree of the network), which usually means a large 
number of them. 
So it is reasonable to expect that the node will be locked in the new strategy 
forever, as it would  require a large number of changes in its neighborhood to 
switch back. 
The great similarity between the simulation results of the original $\infty$T-model 
and the 1T-model supports this intuition 
(see Fig.~\ref{si:fig:01}--\ref{si:fig:04}).

The 1T-model can be exactly calculated for an infinite Bethe lattice, with a recursive 
procedure that we present in the following. 
From now on, let us denote by $X$ one of the options, A or B, 
and by $Y$ the other one, i.e.\ $Y$ is the only element of the set 
$\{\mathrm{A},\mathrm{B}\} - \{X\}$. 
The index of the percolation process with transitions $X \to Y$ is denoted $k_X$.
The fundamental recursive property of the 1T-model is that the time of 
switching of any $X$-node is $1 +$ (the $k_X$-th greater time of switching among 
its $X$-neighbors). 
For example, with a percolation process for transitions $\mathrm{B} \to \mathrm{A}$
with index $k_\mathrm{B}=3$, if a B-node has 6 neighbors that are all 
B-nodes and whose times of switching are $t=$ 2, 2, 3, 5, 7, 7, respectively, 
then the node will switch at time $t=6$. 
Notice that if an $X$-node belongs to a percolating cluster then its time of switching 
is unbounded, because it has $k_X$ or more neighbors that also have unbounded 
switching times.

First, we calculate the switching probabilities of a node, conditioned to the 
event of being connected to a node of the same or the other type. 
Thus, we define $r_X^t$ as the probability of a node being of type $X$ and 
switching at time $t$, conditioned to the event of being connected to an $X$-node 
until time $t-1$, with $t \ge 1$. 
Similarly, we define $s_X^t$ as the probability of a node being of type $X$ and 
switching at time $t$, conditioned to the event of being connected to a $Y$-node, 
i.e.\ of the opposite type, until time $t-1$, with $t \ge 1$. 

Second, we calculate the probabilities of a node being of type $X$ and switching 
at time $t$, which we denote $p_X^t$, given the conditional probabilities of 
switching of its child nodes until time $t-1$, namely $r_X^1,\ldots,r_X^{t-1}$ and $s_Y^1,\ldots,s_Y^{t-1}$.

Third, in the stationary state $X$-nodes will be either nodes that have being 
$X$-nodes since the beginning or initial $Y$-nodes that have switched option. 
Hence the stationary density of B-nodes $x^*$ can be expressed as
\begin{equation}
\label{si:eq:approx-ifc}
x^* = x^0 - \sum_{t=1}^\infty p_\mathrm{B}^t + \sum_{t=1}^\infty p_\mathrm{A}^t .
\end{equation}

To calculate the probabilities 
$\{r_\mathrm{A}^t, s_\mathrm{A}^t, p_\mathrm{A}^t,
   r_\mathrm{B}^t, s_\mathrm{B}^t, p_\mathrm{B}^t\}$ 
we need to consider all the possible configurations of neighborhood of a node. 
To this end, we classify the neighbors at time $t$ into one of these types:\\
$[1]$ Nodes of the same type that switch at time $t-1$. \\
$[2]$ Nodes of the same type that switch at time $t$ or later. \\
$[3]$ Nodes of the same type that switch at time $t-2$ or before. \\
$[4]$ Nodes of the other type that switch at time $t$ or later. \\
$[5]$ Nodes of the other type that switch at time $t-1$ or before. \\
Note that neighbors of type $[1]$ are the ones that trigger switching of the focal 
node at time $t$. For a given configuration, the number of neighbors of each type 
appears as an exponent in the corresponding combinatorial expression.
In Eqs.~(\ref{si:eq:calc-r1})--(\ref{si:eq:calc-P}) below, we use the exponents 
$i, j, k, m, n$ for types $[1]$ to $[5]$, respectively,

In addition, we define  $R_X^0 = S_X^0 = 0$, and $R_X^t = \sum_{\tau=1}^t r_X^\tau$ 
and $S_X^t = \sum_{\tau=1}^t s_X^\tau$, for $t \ge 1$. 
The degree of the Bethe lattice is $z$. The calculation proceeds according to the 
following Eqs.~(\ref{si:eq:calc-r1})--(\ref{si:eq:calc-P})

\begin{equation}
\label{si:eq:calc-r1}
r_X^1 = \sum_{j=0}^{k_X-2} \binom{z-1}{j} (x_X^0)^{j+1} \, (x_Y^0)^{z-1-j} \,,
\end{equation}
\begin{equation}
\label{si:eq:calc-s1}
s_X^1 = \sum_{j=0}^{k_X-1} \binom{z-1}{j} (x_X^0)^{j+1} \, (x_Y^0)^{z-1-j} \,.
\end{equation}

For $t>1$
\begin{equation}
\label{si:eq:calc-rt}
r_X^t = x_X^0 \sum_{(i,j,k,m,n) \in \mathcal{R}} \binom{z-1}{i,j,k,m,n}
 (r_X^{t-1})^i \, (x_X^0-R_X^{t-1})^j \, (R_X^{t-2})^k \,
 (x_Y^0-S_Y^{t-1})^m \, (S_Y^{t-1})^n \,,
\end{equation}
\begin{equation}
\label{si:eq:calc-st}
s_X^t = x_X^0 \sum_{(i,j,k,m,n) \in \mathcal{S}} \binom{z-1}{i,j,k,m,n}
 (r_X^{t-1})^i \, (x_X^0-R_X^{t-1})^j \, (R_X^{t-2})^k \,
 (x_Y^0-S_Y^{t-1})^m \, (S_Y^{t-1})^n \,,
\end{equation}
where the sets of exponents $\mathcal{R}$ and $\mathcal{S}$ are
\begin{align}
\mathcal{R} = \{\; (i,j,k,m,n) \in \{0,1,2,\ldots, z-1\}^5: \quad
& i+j+k+m+n = z-1, \nonumber \\
& i+j+k \ge k_X-1, \nonumber \\
& i+j+n \ge k_X-1, \nonumber \\
\label{si:eq:calc-R}
& j+n \le k_X-2 \,\},
\end{align}
\begin{align}
\mathcal{S} = \{\; (i,j,k,m,n) \in \{0,1,2,\ldots, z-1\}^5: \quad
& i+j+k+m+n = z-1, \nonumber \\
& i+j+k \ge k_X, \nonumber \\
& i+j+n \ge k_X, \nonumber \\
\label{si:eq:calc-S}
& j+n \le k_X-1 \;\}.
\end{align}

\begin{equation}
\label{si:eq:calc-p1}
p_X^1 = \sum_{j=0}^{k_X-1} \binom{z}{j} (x_X^0)^{j+1} \, (x_Y^0)^{z-j} \,.
\end{equation}

For $t > 1$
\begin{equation}
\label{si:eq:calc-pt}
p_X^t = x_X^0 \sum_{(i,j,k,m,n) \in \mathcal{P}} \binom{z}{i,j,k,m,n}
 (r_X^{t-1})^i \, (x_X^0-R_X^{t-1})^j \, (R_X^{t-2})^k \,
 (x_Y^0-S_Y^{t-1})^m \, (S_Y^{t-1})^n \,,
\end{equation}
where the set of exponents $\mathcal{P}$ is
\begin{align}
\mathcal{P} = \{\; (i,j,k,m,n) \in \{0,1,2,\ldots, z\}^5: \quad
& i+j+k+m+n = z, \nonumber \\
& i+j+k \ge k_X, \nonumber \\
& i+j+n \ge k_X, \nonumber \\
\label{si:eq:calc-P}
& j+n \le k_X-1 \;\}.
\end{align}

As the probabilities of switching decay exponentially with time, it is enough to calculate a finite number of terms in the sums of Eq.~(\ref{si:eq:approx-ifc}). 
For the results reported in this work 100 terms were used, which gives excellent 
agreement with the computer simulations (see Figs.~\ref{si:fig:01}--\ref{si:fig:04}).

Notice that Eqs.~(\ref{si:eq:calc-rt})--(\ref{si:eq:calc-S}) 
assume that nodes of the other strategy that have switched (exponent $n$) have done 
so before or at the same time that those of the own strategy (exponent $k$). 
This simplification thus avoids considering the full set of possible histories of 
switchings in times previous to $t-1$. 
This assumption is based on the separation of time scales between the switchings 
of both percolation processes.
For unidirectional percolation, the probability of a node switching at time $t$ 
decreases exponentially with the difference between the initial density and the 
critical percolation density. That is, the nearer the systems starts to the 
percolation threshold the slower the pruning of nodes is 
\cite{grimmett:1999:percolation}. 
As a consequence, with bi-directional percolation, when an option is near its 
percolation threshold the other is well beyond it, so the switching of the 
latter will be exponentially faster, which supports the assumption above. 

It is interesting to point out that the proposed recursive scheme on the switching 
times can be modified to carry out a calculation disregarding interference, thus 
equivalent to Eq.~(\ref{si:eq:approx-no-ifc}). 
The modification consists in setting $s_X^t=0$, for any $t \ge 1$, instead of Eqs.~(\ref{si:eq:calc-s1}) and~(\ref{si:eq:calc-st}). 

Figures~\ref{si:fig:01} and~\ref{si:fig:02} show the results presented in Figs.~2 
and~3 of the main text, but this time including also the simulation results for the 
1T-model and the analytical results obtained with the calculation without interference. 
These figures show the close similarity between the original model and the 1T-model, 
which confirms that most nodes that switch in the original model do so only once. 
They also demonstrate the high accuracy of the proposed approximation, as compared to 
the calculation neglecting interference, which fails to reflect the actual 
percolation thresholds properly. 
Figures~\ref{si:fig:03} and~\ref{si:fig:04} display two additional examples. 

Finally, we want to point out that, strictly speaking, the networks considered in the 
analytical calculations (namely, infinite Bethe lattices) differ from the ones 
used in the computer simulations (finite regular random networks). 
Apart from the different size, in the former case there are no closed paths or 
loops, whereas in the latter there exists a (low) number of them. 
Figures~\ref{si:fig:01}--\ref{si:fig:04} show, however, that these differences 
in network topology do not produce a significative discrepancy between computational 
and analytical results.

\clearpage

\clearpage

\clearpage

\clearpage

\end{document}